\documentclass[prl,twocolumn,showpacs,preprintnumbers,amsmath,amssymb,superscriptaddress]{revtex4-1}

\usepackage{graphicx}
\usepackage{dcolumn}
\usepackage{bm}

\begin{document}\texttt{}


\textbf{Comment on ``Origin of Surface Canting within $\mathrm{Fe}_3\mathrm{O}_4$ Nanoparticles''} \\

In their Letter \cite{kryckaprl2014}, Krycka \textit{et al.} discuss the origin of near-surface spin canting within $\mathrm{Fe}_3\mathrm{O}_4$ nanoparticles by combining magnetic-energy minimization with polarized small-angle neutron scattering (SANS) data. We comment on the SANS data analysis (specifically, Eq.~(1) in \cite{kryckaprl2014}) and on the energy calculations performed in order to find the magnetic ground state of their system.

We start out by commenting the discussion in \cite{kryckaprl2014} regarding the importance of the cross term ($CT$) in Eq.~(1),
\begin{equation}
\label{ct}
CT = - 2 \, | M_{\parallel, X}(\vec{Q}) | \, | M_{\perp, Y = Z}(\vec{Q}) | \, \sin\theta \, \cos^3\theta \, \overline{\cos}(\delta\phi) , \nonumber
\end{equation}
which is used to explain the ``horizontal to vertical suppression'' of the experimental spin-flip data at an applied magnetic field of $1.2$~T (where, according to \cite{kryckaprl2014}, $\overline{\cos}(\delta\phi) = 1$); $M_{\parallel, X}(\vec{Q})$ and $M_{\perp, Y = Z}(\vec{Q})$ denote the (Cartesian) Fourier coefficients of the magnetization and $\theta$ is the angle between the momentum-transfer vector $\vec{Q}$ and the direction of the applied magnetic field $\mathbf{H} \parallel \mathbf{e}_X$ (compare Fig.~1(a) in \cite{kryckaprl2014}). In the Supplemental Material of \cite{kryckaprl2014}, Krycka \textit{et al.} introduce core-shell-type form factors for the functions $M_{\parallel, X}$ and $M_{\perp, Y = Z}$. These single-particle form factors do obviously \textit{not} depend on the orientation (angle $\theta$) of $\vec{Q}$ on the two-dimensional detector, \textit{i.e.}, $M_{\parallel, X} = M_{\parallel, X}(|\vec{Q}|)$ and $M_{\perp, Y = Z} = M_{\perp, Y = Z}(|\vec{Q}|)$. Consequently, the azimuthal average of the $CT$ vanishes, \textit{i.e.}, $\int_0^{2\pi} CT(\theta) d\theta = 0$, demonstrating that the $CT$ does not contribute to the azimuthally-averaged spin-flip SANS cross section or, likewise, to $\pm 10^{\circ}$ sector averages around the horizontal ($\theta = 0^{\circ}$) and vertical ($\theta = 90^{\circ}$) direction. Hence, according to these assumptions made in \cite{kryckaprl2014}, the $CT$ cannot explain the ``horizontal to vertical suppression'' of the spin-flip data, which is, however, a central point of discussion in the Letter. In fact, the main conclusions in \cite{kryckaprl2014} regarding the canting angle of the shell are largely based on the analysis of the horizontal and vertical sector averages.

Furthermore, besides ignoring a term which depends on the polarization of the incident neutrons \cite{blume63}, Eq.~(1) in \cite{kryckaprl2014} assumes that the magnitude-squares of both transversal Fourier coefficients are equal, \textit{i.e.}, $| M_{\perp, Y}(\vec{Q}) |^2 = | M_{\perp, Z}(\vec{Q}) |^2$. These assumptions are not mentioned in \cite{kryckaprl2014}. However, and even more important, the assumption that $| M_{\perp, Y}(\vec{Q}) |^2 = | M_{\perp, Z}(\vec{Q}) |^2$ is questionable, since (for the scattering geometry where $\mathbf{H}$ is perpendicular to the wave vector of the incident neutrons) the magnetodipolar interaction renders both Fourier coefficients different from another: this was shown for \textit{bulk ferromagnets} (two-phase nanocomposites) by means of analytical and numerical micromagnetic simulations \cite{michels2014jmmm,*michels2014review}.

We proceed by commenting on the micromagnetic analysis performed in \cite{kryckaprl2014}. In the first place it should be noted that the spatial discretization used by the authors ($0.05$~nm = $0.5$~$\mathrm{\AA}$) is about $17$ times smaller than the size of the $\mathrm{Fe}_3\mathrm{O}_4$ unit cell ($8.4$~$\mathrm{\AA}$). For such a spatial resolution, the discrete nature of matter should be taken into account when trying to obtain quantitative results, in this particular case, magnetic moments positioned on lattice sites corresponding to the Fe ions. And, even for this (inadequate) spatial discretization, we emphasize that most of the energy expressions used in \cite{kryckaprl2014} for the search of the system's energy minimum are incorrect.

(i) In Eq.~(2) in \cite{kryckaprl2014}, the magnetic anisotropy energy is assumed to be an \textit{uneven} function ($\propto \cos\alpha$; for the definition of $\alpha$ see \cite{kryckaprl2014}). This is inadequate (except for the case of an unidirectional anisotropy, not present here), since, due to fundamental symmetry considerations, magnetic anisotropy energies are \textit{even} functions (\textit{e.g.}, Ref.~\onlinecite{aharonibook}).

(ii) By analyzing the magnetodipolar interaction energy, the authors claim that for a given nanoparticle ``internal dipolar energy is nearly negligible''. This is definitely not true here, because the authors assume that each particle possesses a highly nontrivial magnetization configuration, so that the internal magnetodipolar interaction should play a very important role. Furthermore, the interparticle magnetodipolar interaction is computed incorrectly, because the authors cut-off this interaction after the $18$ nearest neighbors. It is a textbook result that the dipolar interaction is a long-range one \cite{aharonibook}, so that any cut-off of this interaction may lead to arbitrary error and, correspondingly, to unphysical results.

(iii) When computing the anisotropy energy (Eq.~(5) in \cite{kryckaprl2014}) the authors replace the average value of the cosine by the cosine of the average angle, which is clearly an incorrect mathematical operation for any nonlinear function. Moreover, the symmetry of the magnetocrystalline anisotropy is assumed to be uniaxial, although it is well known that $\mathrm{Fe}_3\mathrm{O}_4$ possesses \textit{cubic} anisotropy.

(iv) The exchange energy (Eq.~(6) in \cite{kryckaprl2014}) is proportional to $\cos(T_{d, \mathrm{tilt}})$, where $T_{d, \mathrm{tilt}}$ is defined as the average tilt angle between the $T_d$ Fe sites and the \textit{applied magnetic field}. This is inadequate, since expressions for the exchange interaction (based on the Heisenberg Hamiltonian) depend on the angle between \textit{neighboring magnetic moments} \cite{aharonibook}, and not on the orientation of these moments with respect to the external field.

In conclusion, in view of the substantial criticism raised in this Comment, the conclusions of Krycka \textit{et al.} \cite{kryckaprl2014} regarding the spin structure of $\mathrm{Fe}_3\mathrm{O}_4$ nanoparticles are neither supported by the neutron-data analysis nor by the theoretical considerations.

Financial support from the FNR (FNR/A09/01) and the DFG (BE 2464/10-2) is gratefully acknowledged. 

\vspace{0.5 cm}

\noindent
\normalsize{Andreas Michels,$^{1,*}$ Dirk Honecker,$^2$ Sergey Erokhin,$^3$ and Dmitry Berkov$^3$} \\

\small{$^1$University of Luxembourg, Luxembourg \\
$^2$Institut Laue-Langevin, Grenoble, France \\
$^3$General Numerics Research Lab, Jena, Germany}\\
\small{$^*$Corresponding author: andreas.michels@uni.lu}

\bibliographystyle{apsrev4-1}

\end{document}